\documentclass[conference,a4paper]{IEEEtran}
\IEEEoverridecommandlockouts
\usepackage{cite}
\usepackage{amsmath,amssymb,amsfonts}
\usepackage{algorithmic}
\usepackage{graphicx}
\usepackage{textcomp}
\usepackage{xcolor}
\usepackage[english]{babel}
\usepackage[utf8]{inputenc}
\usepackage{eurosym}
\usepackage{float}
\usepackage{flushend}
\usepackage{setspace}
\usepackage{amssymb}
\usepackage{tikz}
\usetikzlibrary{fit}
\usetikzlibrary{backgrounds}
\usetikzlibrary{positioning,fit,calc}
\usepackage[]{tabularx}
\usepackage{circuitikz}

\usepackage{booktabs,multirow}

\usepackage{moreverb}
\usepackage{enumerate}
\usepackage{textcomp}
\usepackage{graphicx}
\usepackage{amssymb}
\usepackage{amsthm}
\usepackage{amsmath}
\usepackage{multirow}
\usepackage{array}
\usepackage{bm}
\usetikzlibrary{arrows}
\def\BibTeX{{\rm B\kern-.05em{\sc i\kern-.025em b}\kern-.08em
    T\kern-.1667em\lower.7ex\hbox{E}\kern-.125emX}}
\begin{document}

\title{How to overcome the limitations of p-q Theory: Geometric Algebra Power Theory to the rescue}

\author{\IEEEauthorblockN{Francisco G. Montoya}
\IEEEauthorblockA{\textit{Dept. of Engineering} \\
\textit{University of Almeria}\\
Almeria, Spain \\
pagilm@ual.es}
\and
\IEEEauthorblockN{Alfredo Alcayde}
\IEEEauthorblockA{\textit{Dept. of Engineering} \\
\textit{University of Almeria}\\
Almeria, Spain \\
aalcayde@ual.es}
\and
\IEEEauthorblockN{Francisco M. Arrabal-Campos}
\IEEEauthorblockA{\textit{Dept. of Engineering} \\
	\textit{University of Almeria}\\
	Almeria, Spain \\
	fmarrabal@ual.es}
\and
\IEEEauthorblockN{Raúl Baños}
\IEEEauthorblockA{\textit{Dept. of Engineering} \\
	\textit{University of Almeria}\\
	Almeria, Spain \\
	rbanos@ual.es}
}

\maketitle

\begin{abstract}
This paper investigates the recent advances in Geometric Algebra-based power theory (GAPoT) and how this tool provides new insights to solve the flaws of one of the most widespread theory in the time domain, the Instantaneous Reactive Power theory (IRP) and its further enhancements. GAPoT can be applied to single-phase and multi-phase systems to obtain an optimal current decomposition under any distorted voltage source supply and load condition.  This could be the case in microgrids or smart grids. Moreover, it is possible to define different strategies based on instantaneous or averaged quantities depending on whether the voltage supply conditions are sinusoidal and symmetrical or not. Several examples illustrate how GAPoT is able to overcome the limitations of IRP theory.
\end{abstract}


\section{Introduction}

One of the main problems experienced by operators of a smart grid involves the power quality of the grid. Due to the increasing proliferation of non-linear loads and the massive presence of harmonics, it is necessary to establish methods to mitigate the harmful effects of poor power quality on end users. It is therefore imperative to provide appropriate methodologies to assist in maintaining the grid under control for any undesired problem originated by distorted loads. Different techniques have been proposed throughout the last decades \cite{akagi1984instantaneous, depenbrock1993fbd}, with instantaneous reactive power (IRP)  theory being the one that has gained the most acceptance. Formulated in the 80s by Akagi, this theory leads to a current decomposition in the time domain that compensates for the load and preserves its active power consumption. According to its authors, it can be applied under any  circuit conditions and without energy storage requirement \cite{AkagiInstantaneous}. Traditionally, a physical meaning has been attached to current that does not produce a net transfer of energy to the load, known as instantaneous reactive current. This claim has been strongly criticised by several authors \cite{czarnecki2005instantaneous,haley2015limitations} through basic examples that have questioned whether such a process involves an exchange of energy between the source and energy storage elements such as inductors or capacitors. Different studies have shown that strange results are observed in unbalanced load conditions or non-sinusoidal supply, even when linear loads are present \cite{de2005discussion}. 

Further enhancements, based on new transformations and mathematical methods, have also failed to address the shortcomings detected, particularly in the presence of asymmetrical voltages or unbalanced loads \cite{dai2004generalized,peng1996generalized}. In addition, emphasis has been placed on three-phase, three- or four-wire systems, with very minor attention for a generalization to multiple-phase systems \cite{salmeron2009instantaneous}. Nevertheless, the most important aspect is that none of the proposed theories so far have been applied to single-phase systems, so it is not possible to say that there is a generalized theory of power in electrical systems for the time domain.

The goal of this paper is to challenge one of the most widely used theories in the time domain through a comparison with the new GAPoT theory \cite{montoya2020geometric}. Although the former theories were described some decades ago, no significant progress has yet been made in the development of a theory that would coherently explain the exchange of energy between source and load. One of the most likely causes may be that the appropriate mathematical tool has not yet been used.

In the last few years, Geometric Algebra (GA) has demonstrated an inherent ability to deal with multi-component systems in a variety of engineering and scientific fields \cite{montoya2020analysis,cafaro2017geometric}. From quantum physics to robotics, many advances have been achieved through the use of a superior mathematical tool that unifies various techniques used to date such as complex numbers, quaternions, matrices, tensors, etc \cite{hestenes2012clifford}. It has also been a major step  in power systems since it allows to move steadily forward towards a general theory of power that is not limited by the use of complex numbers. As a result, it is now possible to define a power that is conservative and takes into account the interactions between voltage and current harmonics of different frequencies. This has not been possible before. 
Through several  examples already proposed in the literature (specifically designed to demonstrate the shortcomings of IRP theory), it will be shown how GAPoT can give a feasible and elegant solution, so that coherent results are obtained fulfilling the expected physical principles.

\section{Power theory approaches}

\subsection{Instantaneous reactive power theory}
\begin{figure} % width of left subfigure
	\centering
	\tikzset{every picture/.style={line width=0.75pt}} %set default line width to 0.75pt        
	\begin{tikzpicture}[x=0.6pt,y=0.6pt,yscale=-1,xscale=1]
	%uncomment if require: \path (0,446.1999969482422); %set diagram left start at 0, and has height of 446.1999969482422
	
	%Rounded Rect [id:dp4780074834509005] 
	\draw  [fill={rgb, 255:red, 238; green, 229; blue, 229 }  ,fill opacity=0.38 ] (100,156.13) .. controls (100,141.88) and (111.55,130.33) .. (125.8,130.33) -- (203.2,130.33) .. controls (217.45,130.33) and (229,141.88) .. (229,156.13) -- (229,244.4) .. controls (229,258.65) and (217.45,270.2) .. (203.2,270.2) -- (125.8,270.2) .. controls (111.55,270.2) and (100,258.65) .. (100,244.4) -- cycle ;
	%Straight Lines [id:da28286023504566815] 
	\draw [line width=1.5]    (229,162.13) -- (382.67,162.13) ;

	%Straight Lines [id:da01650347726715462] 
	\draw [line width=1.5]    (229.67,199.13) -- (383.33,199.13) ;

	%Straight Lines [id:da6653434815696049] 
	\draw [line width=1.5]    (228.67,239.53) -- (382.33,239.53) ;

	%Rounded Rect [id:dp3977749994943989] 
	\draw  [fill={rgb, 255:red, 238; green, 232; blue, 232 }  ,fill opacity=0.3 ] (382.33,156.13) .. controls (382.33,141.88) and (393.88,130.33) .. (408.13,130.33) -- (485.53,130.33) .. controls (499.78,130.33) and (511.33,141.88) .. (511.33,156.13) -- (511.33,243) .. controls (511.33,257.25) and (499.78,268.8) .. (485.53,268.8) -- (408.13,268.8) .. controls (393.88,268.8) and (382.33,257.25) .. (382.33,243) -- cycle ;
	%Straight Lines [id:da16546759790135335] 
	\draw    (339.7,162.2) -- (363,162.02) ;
	\draw [shift={(365,162)}, rotate = 539.55] [fill={rgb, 255:red, 0; green, 0; blue, 0 }  ][line width=0.75]  [draw opacity=0] (8.93,-4.29) -- (0,0) -- (8.93,4.29) -- cycle    ;
	
	%Straight Lines [id:da9228558239915912] 
	\draw    (339.3,199) -- (362.6,198.82) ;
	\draw [shift={(364.6,198.8)}, rotate = 539.55] [fill={rgb, 255:red, 0; green, 0; blue, 0 }  ][line width=0.75]  [draw opacity=0] (8.93,-4.29) -- (0,0) -- (8.93,4.29) -- cycle    ;
	
	%Straight Lines [id:da1187517392405808] 
	\draw    (340.2,239.7) -- (363.5,239.52) ;
	\draw [shift={(365.5,239.5)}, rotate = 539.55] [fill={rgb, 255:red, 0; green, 0; blue, 0 }  ][line width=0.75]  [draw opacity=0] (8.93,-4.29) -- (0,0) -- (8.93,4.29) -- cycle    ;
	
	%Shape: Circle [id:dp14910226292447448] 
	\draw  [fill={rgb, 255:red, 255; green, 255; blue, 255 }  ,fill opacity=1 ] (237.5,165.04) .. controls (235.8,163.3) and (235.84,160.51) .. (237.58,158.81) .. controls (239.32,157.11) and (242.11,157.14) .. (243.8,158.88) .. controls (245.5,160.62) and (245.47,163.41) .. (243.73,165.11) .. controls (241.99,166.81) and (239.2,166.78) .. (237.5,165.04) -- cycle ;
	%Shape: Circle [id:dp6653234598411544] 
	\draw  [fill={rgb, 255:red, 255; green, 255; blue, 255 }  ,fill opacity=1 ] (237.5,243.04) .. controls (235.8,241.3) and (235.84,238.51) .. (237.58,236.81) .. controls (239.32,235.11) and (242.11,235.14) .. (243.8,236.88) .. controls (245.5,238.62) and (245.47,241.41) .. (243.73,243.11) .. controls (241.99,244.81) and (239.2,244.78) .. (237.5,243.04) -- cycle ;
	%Shape: Circle [id:dp44733264311243004] 
	\draw  [fill={rgb, 255:red, 255; green, 255; blue, 255 }  ,fill opacity=1 ] (237.5,202.54) .. controls (235.8,200.8) and (235.84,198.01) .. (237.58,196.31) .. controls (239.32,194.61) and (242.11,194.64) .. (243.8,196.38) .. controls (245.5,198.12) and (245.47,200.91) .. (243.73,202.61) .. controls (241.99,204.31) and (239.2,204.28) .. (237.5,202.54) -- cycle ;
	%Shape: Circle [id:dp2688418516872404] 
	\draw  [fill={rgb, 255:red, 255; green, 255; blue, 255 }  ,fill opacity=1 ] (152,174.73) .. controls (152,168.5) and (157.05,163.46) .. (163.27,163.46) .. controls (169.5,163.46) and (174.54,168.5) .. (174.54,174.73) .. controls (174.54,180.95) and (169.5,186) .. (163.27,186) .. controls (157.05,186) and (152,180.95) .. (152,174.73) -- cycle ;
	%Shape: Circle [id:dp5512711868999121] 
	\draw  [fill={rgb, 255:red, 255; green, 255; blue, 255 }  ,fill opacity=1 ] (178.5,220.23) .. controls (178.5,214) and (183.55,208.96) .. (189.77,208.96) .. controls (196,208.96) and (201.04,214) .. (201.04,220.23) .. controls (201.04,226.45) and (196,231.5) .. (189.77,231.5) .. controls (183.55,231.5) and (178.5,226.45) .. (178.5,220.23) -- cycle ;
	%Shape: Circle [id:dp35363260965571897] 
	\draw  [fill={rgb, 255:red, 255; green, 255; blue, 255 }  ,fill opacity=1 ] (123,220.23) .. controls (123,214) and (128.05,208.96) .. (134.27,208.96) .. controls (140.5,208.96) and (145.54,214) .. (145.54,220.23) .. controls (145.54,226.45) and (140.5,231.5) .. (134.27,231.5) .. controls (128.05,231.5) and (123,226.45) .. (123,220.23) -- cycle ;
	%Straight Lines [id:da9429570886492475] 
	\draw    (163.27,186) -- (163.5,205.46) ;

	%Straight Lines [id:da87455960366007] 
	\draw    (163.5,205.46) -- (145,215.96) ;

	%Straight Lines [id:da3250735712945574] 
	\draw    (163.5,205.46) -- (179.5,215.46) ;

	%Curve Lines [id:da24883277937851012] 
	\draw    (156.28,175.88) .. controls (162.68,161.88) and (162.28,187.08) .. (169.88,173.48) ;

	%Curve Lines [id:da36546279293952444] 
	\draw    (183.08,221.48) .. controls (189.48,207.48) and (189.08,232.68) .. (196.68,219.08) ;

	%Curve Lines [id:da013851789881527132] 
	\draw    (127.48,221.08) .. controls (133.88,207.08) and (133.48,232.28) .. (141.08,218.68) ;

	%Flowchart: Process [id:dp20636044072987603] 
	\draw  [fill={rgb, 255:red, 255; green, 255; blue, 255 }  ,fill opacity=1 ] (442.7,167.7) -- (450.7,167.7) -- (450.7,189.3) -- (442.7,189.3) -- cycle ;
	%Flowchart: Process [id:dp15780770095359253] 
	\draw  [fill={rgb, 255:red, 255; green, 255; blue, 255 }  ,fill opacity=1 ] (486.45,228.04) -- (482.45,234.96) -- (463.75,224.16) -- (467.75,217.24) -- cycle ;
	%Flowchart: Process [id:dp17027095359237765] 
	\draw  [fill={rgb, 255:red, 255; green, 255; blue, 255 }  ,fill opacity=1 ] (426.45,216.44) -- (430.45,223.36) -- (411.75,234.16) -- (407.75,227.24) -- cycle ;
	%Straight Lines [id:da4921196628946476] 
	\draw    (447.27,190) -- (447.5,209.46) ;

	%Straight Lines [id:da6453335886757787] 
	\draw    (447.5,209.46) -- (429,219.96) ;

	%Straight Lines [id:da4988148992363355] 
	\draw    (447.5,209.46) -- (465.73,220.62) ;

	%Straight Lines [id:da6020319405389247] 
	\draw    (409.5,230.66) -- (392.93,239.42) ;

	%Straight Lines [id:da6961040076165983] 
	\draw    (499.73,240.62) -- (484.93,232.22) ;

	%Straight Lines [id:da4789222010393117] 
	\draw    (446.87,148.4) -- (447.1,167.86) ;

	%Flowchart: Process [id:dp9633957840526868] 
	\draw  [fill={rgb, 255:red, 255; green, 255; blue, 255 }  ,fill opacity=1 ][dash pattern={on 4.5pt off 4.5pt}] (422.08,182.6) -- (428.92,186.75) -- (417.72,205.22) -- (410.88,201.07) -- cycle ;
	%Straight Lines [id:da3762253537332718] 
	\draw  [dash pattern={on 4.5pt off 4.5pt}]  (414.13,203.62) -- (392.93,239.42) ;

	%Straight Lines [id:da12033250139560336] 
	\draw  [dash pattern={on 4.5pt off 4.5pt}]  (446.87,148.4) -- (425.67,184.2) ;

	%Flowchart: Process [id:dp9573417924460146] 
	\draw  [fill={rgb, 255:red, 255; green, 255; blue, 255 }  ,fill opacity=1 ][dash pattern={on 4.5pt off 4.5pt}] (464.36,187.61) -- (471.36,183.73) -- (481.84,202.61) -- (474.85,206.49) -- cycle ;
	%Straight Lines [id:da8144739586851588] 
	\draw  [dash pattern={on 4.5pt off 4.5pt}]  (478.67,204.93) -- (499.73,240.62) ;

	%Straight Lines [id:da5352716057434048] 
	\draw  [dash pattern={on 4.5pt off 4.5pt}]  (446.97,149.12) -- (467.53,185.28) ;

	%Flowchart: Process [id:dp13576509412684734] 
	\draw  [fill={rgb, 255:red, 255; green, 255; blue, 255 }  ,fill opacity=1 ][dash pattern={on 4.5pt off 4.5pt}] (457.53,236.38) -- (457.48,244.38) -- (435.88,244.24) -- (435.93,236.24) -- cycle ;
	%Straight Lines [id:da1367976131016413] 
	\draw  [dash pattern={on 4.5pt off 4.5pt}]  (435.41,240.34) -- (392.93,239.42) ;

	%Straight Lines [id:da4744840230330516] 
	\draw  [dash pattern={on 4.5pt off 4.5pt}]  (499.73,240.62) -- (458,240.28) ;

	%Straight Lines [id:da11191781331001338] 
	\draw [line width=2.25]    (160,292) -- (450,292) ;
	\draw [line width=2.25]    (160,292) -- (160,270) ;
	\draw [line width=2.25]    (450,292) -- (450,269) ;
	
	%Straight Lines [id:da2363772277999201] 
	\draw    (270,170) -- (270,288) ;
	\draw [shift={(270,290)}, rotate = 270] [fill={rgb, 255:red, 0; green, 0; blue, 0 }  ][line width=0.75]  [draw opacity=0] (8.93,-4.29) -- (0,0) -- (8.93,4.29) -- cycle    ;
	
	%Straight Lines [id:da39027724404200703] 
	\draw    (280,210) -- (280,288) ;
	\draw [shift={(280,290)}, rotate = 270] [fill={rgb, 255:red, 0; green, 0; blue, 0 }  ][line width=0.75]  [draw opacity=0] (8.93,-4.29) -- (0,0) -- (8.93,4.29) -- cycle    ;
	
	%Straight Lines [id:da8032215277379307] 
	\draw    (290,250) -- (290,288) ;
	\draw [shift={(290,290)}, rotate = 270] [fill={rgb, 255:red, 0; green, 0; blue, 0 }  ][line width=0.75]  [draw opacity=0] (8.93,-4.29) -- (0,0) -- (8.93,4.29) -- cycle    ;

	% Text Node
	\draw (161.5,111) node  [align=left] {Power Supply};
	% Text Node
	\draw (449,111) node  [align=left] {Load};
	% Text Node
	\draw (351,144.83) node   {$\bm{i}_{R}$};
	% Text Node
	\draw (352,182.83) node   {$\bm{i}_{S}$};
	% Text Node
	\draw (352,225.83) node   {$\bm{i}_{T}$};
	% Text Node
	\draw (241,144.83) node   {$\bm{R}$};
	% Text Node
	\draw (241,180.83) node   {$\bm{S}$};
	% Text Node
	\draw (240.5,219.83) node   {$\bm{T}$};
	% Text Node
	\draw (249,280) node   {$\bm{N}$};
	% Text Node
	\draw (290,172.5) node   {$\bm{u}_{RN}$};
	% Text Node
	\draw (300,212.5) node   {$\bm{u}_{SN}$};
	% Text Node
	\draw (311,252.5) node   {$\bm{u}_{TN}$};
	
	\end{tikzpicture}
	\caption{Three-phase, four wire circuit}
	\label{fig:threewire_system}
\end{figure}
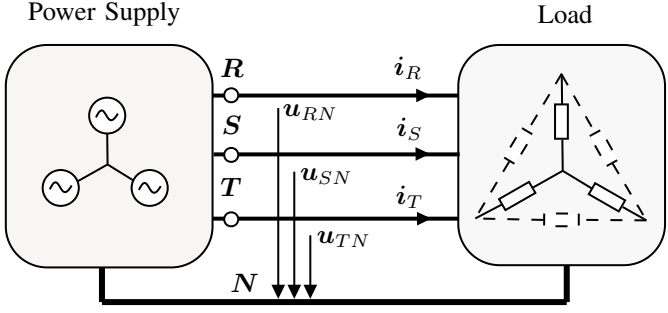

The IRP theory form analyzed in this work is the one presented in \cite{peng1996generalized} and supported by \cite{dai2004generalized}. It is called cross-vector (CV) generalized theory because it is based on the use of the vector product. In \cite{salmeron2009instantaneous} it is extended to multiple phases through the tensor product. The CV theory is established for four-wire systems like the one in Figure \ref{fig:threewire_system}, where voltage and current are defined as,

\begin{equation}
	\begin{aligned}
	\bm{u}(t)\
	=
	\begin{bmatrix}
	\bm{u}_R(t) \\ \bm{u}_S(t) \\ \bm{u}_T(t)
	\end{bmatrix},\;\;
	\bm{i}(t)
	=
	\begin{bmatrix}
	\bm{i}_R(t) \\ \bm{i}_S(t) \\\bm{i}_T(t)
	\end{bmatrix}
	\end{aligned}
\end{equation}

The instantaneous power is obtained as

\begin{equation}
	p(t)=\bm{u}^T\bm{i}=u_Ri_R+u_Si_S+u_Ti_T
\end{equation}

\noindent where the time symbol $(t)$ for voltage and current has been omitted for convenience. The concept of instantaneous reactive power is introduced mathematically through the vector product of voltage and current

\begin{equation}
\begin{aligned}
\bm{q}(t)\
= \bm{u}\times \bm{i}=
\begin{bmatrix}
u_Si_T-u_Ti_S \\ u_Ti_R-u_Ri_T \\ u_Ri_S-u_Si_R
\end{bmatrix}
\end{aligned}
\label{eq:q_instant}
\end{equation}

Contrary to instantaneous power, this quantity is a vector. The norm is defined as

\begin{equation}
	q(t)=\sqrt{\bm{q}^T\bm{q}}
\end{equation}

Based on this approach, the IRP theory defines a current decomposition as follows

\begin{equation}
\begin{aligned}
\bm{i}_p\
&= 
\begin{bmatrix}
i_{R_p} \\ i_{S_p} \\ i_{T_p}
\end{bmatrix}
=\frac{p(t)}{\bm{u}^T\bm{u}}\bm{u}=G_p(t)\bm{u}
\\
\bm{i}_q\
&=
\begin{bmatrix}
i_{R_q} \\ i_{S_q} \\ i_{T_q}
\end{bmatrix}
=\frac{\bm{q}\times \bm{u}}{\bm{u}^T\bm{u}}\bm{u}
\end{aligned}
\end{equation}

\noindent so that $\bm{i}=\bm{i}_p+\bm{i}_q$. It follows from the definition of $\bm{i}_p$ and $\bm{i}_q$ that both vectors are orthogonal, i.e.

\begin{equation}
	\bm{i}_p^T\bm{i}_q=\bm{i}_q^T\bm{i}_p=0
\end{equation}

The current $\bm{i}_q$ can be compensated without energy storage so that only $\bm{i}_p$ remains after compensation. This results in a reduction in the original current $\bm{i}$. The norm of the total current can be calculated as

\begin{equation}
	\|\bm{i}\|^2=\bm{i}^T\bm{i}=\bm{i}_p^T\bm{i}_p+\bm{i}_q^T\bm{i}_q
\end{equation}

Despite the benefit of reducing the RMS current for the same active power, severe limitations have been described in the literature \cite{czarnecki2005instantaneous}. Perhaps one of the most significant is that this theory cannot be applied to single phase circuits, since as defined in (\ref{eq:q_instant}), its value is always zero for such systems.

\subsection{GAPoT theory}
Recently,  GAPoT theory has been developed for multiphase systems in the time domain \cite{montoya2020geometric}.  The apparent power has been redefined through the use of geometric algebra. In addition, the Hilbert transform (HT) has been used for voltage and current definitions. The use of these mathematical tools leads to a very compact formulation. GAPoT can be applied in the most general sense, including single and multiphase systems, linear and non-linear circuits, sinusoidal and non-sinusoidal power supply, symmetrical or asymmetrical power supply, balanced and unbalanced loads. It can also be used according to the two key criteria in time domain power theory: averaged or instantaneous quantities. The former is an original contribution of this theory, while the latter is a natural extension of the existing formulation. 

GAPoT relies on the use of an orthonormal base $\bm{\sigma}=\{ \bm{\sigma}_1,\bm{\sigma}_2,\ldots,\bm{\sigma}_n\}$ defined for a vector space in $\mathcal{R}^n$. Then, it is possible to establish a new geometric vector space $\mathcal{G}^n$ with a bilinear form. Under these assumptions, a vector can be represented as:
\begin{equation}
\bm{v}=\sum_{n}v_n\bm{\sigma}_n=v_1\bm{\sigma}_1+\ldots+v_n\bm{\sigma}_n
\end{equation}
In this new space, the geometric product between two vectors ($\bm{u}$ and $\bm{v}$) can be defined as:
\begin{equation}
\bm{M}=\bm{uv}=\bm{u}\cdot\bm{v}+\bm{u}\wedge\bm{v}
\end{equation}
which can be seen as the sum of the traditional scalar or inner product plus the so-called wedge or Grassmann product. The latter fulfils the anticommutativity property:
\begin{equation}
\bm{u}\wedge\bm{v}=-\bm{v}\wedge\bm{u}
\label{ec.bivector}
\end{equation}
The above entity is commonly known as {\em bivector} and is a new object not found previously in linear algebra.  

For a multiphase system, the phase voltages and line currents can be defined by the following arrays

\begin{equation}
	\begin{aligned}
	\Vec{u}(t)&=\left[u_1, u_2, \: \ldots \:, u_n\right] \\
	\Vec{i}(t)&=\left[i_1, i_2, \: \ldots \:, i_n\right]
	\end{aligned}
\end{equation}

\noindent so that, according to GAPoT, they can be transferred to the geometric domain as

\begin{equation}
\setlength{\arraycolsep}{0pt}
\begin{array}{ r *{5}{ >{{}}c<{{}} r } }
\bm{u} &=&
u_1\bm{\sigma}_1 &+&
\bm{\mathcal{H}}\left[u_1\right]\bm{\sigma}_2 &+&
\cdots &+&
u_n\bm{\sigma}_{2n-1} &+&
\bm{\mathcal{H}}\left[u_n\right]\bm{\sigma}_{2n}
\\[1ex]
\bm{i} &=&
i_1\bm{\sigma}_1 &+&
\bm{\mathcal{H}}\left[i_1\right]\bm{\sigma}_2 &+&
\cdots &+&
i_n\bm{\sigma}_{2n-1} &+&
\bm{\mathcal{H}}\left[i_n\right]\bm{\sigma}_{2n}
\end{array}
\label{eq:GAPoT_transform} 
\end{equation}

\noindent where the operator $\bm{\mathcal{H}}$ refers to the Hilbert transform, defined as in \cite{bracewell1986fourier}

\begin{equation}
\bm{\bm{\bm{\mathcal{H}}}}\left[u(t)\right]=\frac{1}{\pi}PV\int_{-\infty}^{+\infty}-\frac{u(\tau)}{t-\tau}d\tau
\end{equation}

It should be noted that HT is only required for averaged quantities in multiphase or single-phase systems. It can be omitted for the study of instantaneous multiphase systems. Single-phase systems cannot be compensated instantaneously. Instantaneous geometric power is the product of the voltage vector $\bm{u}$ and the current vector $\bm{i}$

\begin{equation}
\bm{M}=\bm{ui}=\bm{u}\cdot\bm{i}+\bm{u}\wedge\bm{i}=M_p+\bm{M}_q
\label{eq:geom_power_time}
\end{equation}

\noindent which consists of a scalar part $M_p=\bm{u}\cdot\bm{i}$ and a bivector part $\bm{M}_q=\bm{u\wedge\bm{i}}$. $M_p$ is the  \textit{parallel geometric power} and includes the instantaneous active power $p(t)$. $\bm{M_q}$ is \textit{quadrature geometric power} and it comprises the well-known \textit{instantaneous reactive power} in the CV theory.

It follows from (\ref{eq:geom_power_time}) that the current can be cleared from the equation by left-multiplying by the inverse of the voltage 

\begin{equation}
\begin{aligned}
\bm{i}&=\bm{u^{-1}M}=\frac{\bm{u}}{\|\bm{u}\|^2}\bm{M}=\frac{\bm{u}}{\|\bm{u}\|^2}\left(M_p+\bm{M_q}\right) \\
&=\frac{\bm{u}}{\|\bm{u}\|^2}M_p+\frac{\bm{u}}{\|\bm{u}\|^2}\bm{M_q}=\bm{i_p}+\bm{i_q}
\end{aligned} 
\label{eq:current_decomposition_initial}
\end{equation}

\noindent where $\bm{u}^{-1}=\frac{\bm{u}}{\|\bm{u}\|^2}$. The current decomposition in (\ref{eq:current_decomposition_initial}) occurs naturally, and this is an inherent advantage of the proposed theory. It can be readily demonstrated that the pairs $\bm{i}_p$ - $\bm{i}_q$ and $M_p$ - $\bm{M}_q$ are orthogonal \cite{montoya2020geometric}.
The Fryze current can be also included in the GAPoT theory as

\begin{equation}
\bm{i_F}=\frac{\bar{M}_p}{\|\bm{\bar{\bm{u}}}\|^2}\bm{u}
\end{equation}

\noindent where $\bar{M}_p$ is the mean value of the geometric parallel power and $\|\bar{\bm{u}}\|$ is the RMS value of the geometric voltage. It can be readily demonstrated that $\bar{M}_p=2P$, where $P$ is the active power. The Budeanu reactive current is also defined as

\begin{equation}
\bm{i_B}=\frac{\bar{M}_q}{\|\bm{\bar{\bm{u}}}\|^2}\bm{\mathcal{H}}\left[\bm{u}\right]
\end{equation}

\noindent where $\bar{M}_q$ is the mean value of the quadrature geometric power. The complete current decomposition is

\begin{equation}
\bm{i}=\bm{i_p}+\bm{i_q}=\bm{i_F}+\bm{i_f}+\bm{i_B}+\bm{i_b}
\label{eq:todas_corrientes}
\end{equation}

\noindent where $\bm{i_f}$ is the Fryze complementary current required to conform the parallel current. Similarly, $\bm{i_b}$ is the Budeanu complementary current required to conform the quadrature current.

Through several examples, the new theory will be compared with the CV theory. Results will be presented that are in accordance with the physical and engineering principles expected when using the averaging strategy. Also, a single phase circuit will be solved to show the benefits of the proposed theory.

\section{Examples}
%An unbalanced three-phase circuit and a single-phase circuit will be solved by GAPoT to verify the capability of the method.
\subsection{Illustration 1}

\begin{figure}
	\centering
	\includegraphics[width=0.3\textwidth]{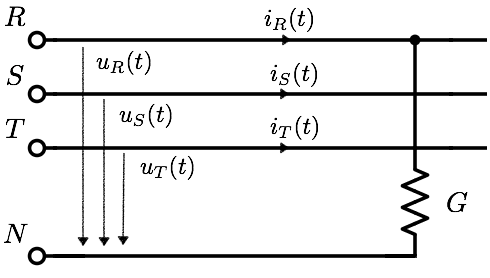}
	\caption{Unbalanced three-phase, four wire circuit}
	\label{fig:three_pase}
\end{figure}

The unbalanced circuit in figure \ref{fig:three_pase} has been studied in \cite{haley2015limitations}. The voltage source is 

\begin{equation*}
\begin{array}{l}
u_R(t)=\sqrt{2}U\cos \omega t \\ u_S(t)=\sqrt{2}U\cos\left(\omega t -120 \right)\\ u_T(t)=\sqrt{2}U\cos\left(\omega t +120 \right)
\end{array}
\end{equation*}

\noindent and the current is

\begin{equation*}
i_R(t)=\sqrt{2}GU\cos \omega t
\end{equation*}

Compensation by CV theory produces conflicting results for the power factor \cite{haley2015limitations}. Also, the compensated current is asymmetrical and distorted in spite of being a purely linear resistive circuit fed by a symmetrical and sinusoidal voltage. In particular, the current after compensation proposed by the CV theory is 

\begin{equation*}
	\bm{i}_p^{CV}=\frac{G_R+G_R\cos(2\omega t)}{3}\bm{u}
\end{equation*}

\noindent which contains third-order harmonics and negative sequence components. In addition, it is also found that the power factor is smaller than unity  after compensation.

Applying the instantaneous strategy in GAPoT leads to the same results as the CV theory. However, the averaging strategy leads to different results, according to the expected physical principles of the problem. 

The geometric instantaneous voltage and current vector can be derived according to (\ref{eq:GAPoT_transform}) 

\begin{equation*}
\begin{aligned}
&\bm{u} =\sqrt{2}U[\cos \omega t\bm{\sigma}_1-\sin \omega t\bm{\sigma}_2 +\cos \left(\omega t-120\right)\bm{\sigma}_3\\
 &-\sin \left(\omega t-120\right) \bm{\sigma}_4 +\cos \left(\omega t+120\right)\bm{\sigma}_5 -\sin \left(\omega t+120\right) \bm{\sigma}_6]
\\[1ex]
&\bm{i} =\sqrt{2}GU[\cos \omega t\bm{\sigma}_1 -\sin \omega t\bm{\sigma}_2]
\end{aligned}
\end{equation*}

Their product yields the geometric apparent power

\begin{equation*}
\begin{aligned}
\bm{M}&=2GU^2\left[1 -\right.\\
&\left.-\cos \omega t \cos \left(\omega t - 120\right)\bm{\sigma}_{13} 
+\cos \omega t \sin \left(\omega t - 120\right)\bm{\sigma}_{14} \right.\\
&\left.-\cos \omega t \cos \left(\omega t + 120\right)\bm{\sigma}_{15} 
 +\cos \omega t \sin \left(\omega t + 120\right)\bm{\sigma}_{16}\right.\\
&\left.+\sin \omega t \cos \left(\omega t - 120\right)\bm{\sigma}_{23} 
-\sin \omega t \sin \left(\omega t - 120\right)\bm{\sigma}_{24}  \right.\\
&\left.+\sin \omega t \cos \left(\omega t + 120\right)\bm{\sigma}_{25}
-\sin \omega t \sin \left(\omega t + 120\right)\bm{\sigma}_{26}    \right]
\end{aligned}
\end{equation*}

In the above expression, the parallel power is constant with a value of $M_p=2GU^2$. This is consistent with expectations, since there is a resistive circuit with active power  $P=GU^2=\bar{M}_p/2$. The rest of terms are bivector elements conforming the quadrature power and are related to the load imbalance. Note that there are no terms  $\bm{\sigma}_{12}$, $\bm{\sigma}_{34}$, $\bm{\sigma}_{56}$, so there is no presence of reactive power in the Budeanu sense as  expected in the absence of inductive or capacitive elements.

Once the geometric power has been found, the current decomposition can be calculated according to (\ref{eq:current_decomposition_initial})-(\ref{eq:todas_corrientes}), bearing in mind that $\|\bm{u}\|^2=6U^2$

\begin{equation*}
\begin{aligned}
\bm{i}_p&=\frac{\bm{u}}{\|\bm{u}\|^2}M_p=\frac{G}{3}\bm{u}=\sqrt{2}\frac{GU}{3}[\cos \omega t\bm{\sigma}_1 -\sin \omega t\bm{\sigma}_2 \\
&+\cos \left(\omega t-120\right)\bm{\sigma}_3 -\sin \left(\omega t-120\right) \bm{\sigma}_4 \\
&+\cos \left(\omega t+120\right)\bm{\sigma}_5 -\sin \left(\omega t+120\right) \bm{\sigma}_6]\\[1ex]
\bm{i}_q&=\frac{\bm{u}}{\|\bm{u}\|^2}\bm{M_q}=\bm{i}-\bm{i}_p=\sqrt{2}\frac{GU}{3}[2\cos \omega t\bm{\sigma}_1 -2\sin \omega t\bm{\sigma}_2 \\
&+\cos \left(\omega t-120\right)\bm{\sigma}_3 -\sin \left(\omega t-120\right) \bm{\sigma}_4 \\
&+\cos \left(\omega t+120\right)\bm{\sigma}_5 -\sin \left(\omega t+120\right) \bm{\sigma}_6]\\[1ex]
\end{aligned}
\end{equation*}

%		\bm{i}_F&=\frac{\bar{M}_p}{\|\bm{\bar{\bm{u}}}\|^2}\bm{u}\\
%		\bm{i}_f&=\bm{i}_p-\bm{i}_F\\
%		\bm{i}_B&=\frac{\bar{M}_q}{\|\bm{\bar{\bm{u}}}\|^2}\bm{\mathcal{H}}\left[\bm{u}\right]\\
%		\bm{i}_b&=\bm{i}_q-\bm{i}_B

Furthermore, in this example, the Fryze current matches the parallel current, that is, $\bm{i}_p=\bm{i}_F$, and therefore, $\bm{i}_f=0$. There is also no reactive Budeanu current since  $\bar{M}_q=0$, so $\bm{i}_B=0$. Thus, the current $\bm{i}_b=\bm{i}_q$, meaning that it contains all the asymmetry components, i.e. the zero-sequence current $\bm{i}_0$ and the inverse-sequence current $\bm{i}_{-}$

\begin{equation*}
\begin{aligned}
&\bm{i}_q=\bm{i}_b=\bm{i}_0+\bm{i}_{-}=\sqrt{2}\frac{GU}{3}[\cos \omega t\bm{\sigma}_1 -\sin \omega t\bm{\sigma}_2+\cos \omega t\bm{\sigma}_3 \\
	&-\sin \omega t\bm{\sigma}_4+\cos \omega t\bm{\sigma}_5 -\sin \omega t\bm{\sigma}_6]\\
&+\sqrt{2}\frac{GU}{3}[\cos \omega t\bm{\sigma}_1 -\sin \omega t\bm{\sigma}_2 +\cos \left(\omega t+120\right)\bm{\sigma}_3 \\
&-\sin \left(\omega t+120\right) \bm{\sigma}_4 +\cos \left(\omega t-120\right)\bm{\sigma}_5 -\sin \left(\omega t-120\right) \bm{\sigma}_6]
\end{aligned}
\end{equation*}

\begin{figure}
	\centering
	\includegraphics[width=0.5\textwidth]{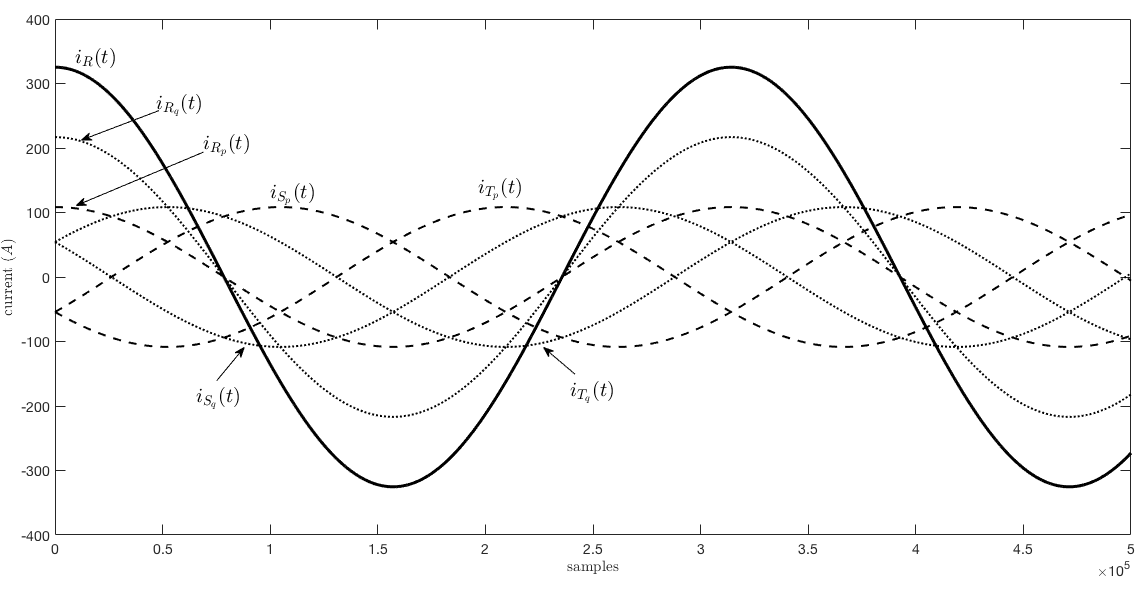}
	\caption{Currents decomposition for three-phase circuit}
	\label{fig:three_pase_currents}
\end{figure}

The time domain currents are recovered by applying 

\begin{equation*}
i(t) = \sum_{k=1}^{n}[\bm{i}]_{2k-1}
\label{eq:projection_currents}
\end{equation*}
\noindent where $[\cdot]_k$ refers to the $k$-th component  of the current geometric vector $\bm{i}$.
Figure \ref{fig:three_pase_currents} depicts the current decomposition for this problem assuming $U=230$, $\omega=1$ and $G=1$.

The power factor of the circuit can be found using 

\begin{equation}
	pf=\frac{\bar{M}_p}{\|\bm{M}\|}=\frac{\bar{M}_p}{\|\bm{u}\|\|\bm{i}\|} 
\end{equation}

\noindent so that, for the uncompensated circuit, the result is

\begin{equation*}
pf_{\text{orig}}=\frac{\bar{M}_p}{\|\bm{M}_{\text{orig}}\|}=\frac{2GU^2}{\sqrt{6}U\sqrt{2}GU}=\frac{1}{\sqrt{3}}=0.577
\end{equation*}

For the compensated circuit, the power factor achieved is the unity as expected,

\begin{equation*}
pf_{\text{final}}=\frac{\bar{M}_p}{\|\bm{M}_{\text{final}}\|}=\frac{2GU^2}{\sqrt{6}U\sqrt{\frac{2}{3}}GU}=1
\end{equation*}

\subsection{Illustration 2}

\begin{figure}
	\centering
	\begin{circuitikz}[scale=1.5] \draw
		(0,0) to[sV, v<=$u(t)$] (0,1) 
%		(3,0) -- (0,0)
		(0,1) to [R, label=${R=1}$, i>^=$i$] (2,1)
		to [L, label=${L=\dfrac{1}{2}}$] (3,1)
		to [C, label=${C=1}$] (5,1)
		(5,1) -- (5,0) -- (0,0)
		%	(1.5,4) to[R, european resistor, label=$Y_{cp}$, i>^=$i_{cp}$] (1.5,0)
		;						
		%	\draw (-0.3,2.3) node{$+$};						
	\end{circuitikz}
	\caption{Single phase RLC circuit}
	\label{fig:RLC}
\end{figure}
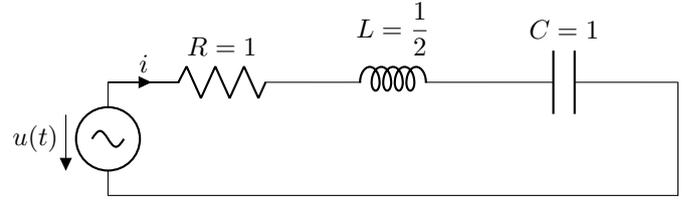

Figure \ref{fig:RLC} shows a simple $RLC$ single-phase circuit. According to CV theory, no meaningful results can be drawn because the instantaneous reactive power is always a zero vector according to (\ref{eq:q_instant}). Consider a non-sinusoidal supply $u(t)=100\sqrt{2}\left(\cos t + \cos 3t\right)$. Applying (\ref{eq:GAPoT_transform}), the geometric voltage and current are

\begin{equation}
\begin{aligned}
\bm{u}&=100\sqrt{2}[(\cos t + \cos 3t)\bm{\sigma}_1+ (-\sin t -\sin 3t)\bm{\sigma}_2]\\
\bm{i}&=\sqrt{2}[(80\cos t - 40\sin t+42.35\cos 3t +49.41\sin 3t)\bm{\sigma}_1\\
&+ (-80\sin t - 40\cos t-42.35\sin 3t +49.41\cos 3t)\bm{\sigma}_2]
\end{aligned}
\end{equation}

\begin{table*}[ht!]
	\centering
	\begin{tabular}{@{}cllr@{}}
		\toprule
		&  \multicolumn{2}{c}{\textbf{vector}}&  \\ \cmidrule{2-3}
		& \multicolumn{1}{c}{$\bm{\sigma}_1$} & \multicolumn{1}{c}{$\bm{\sigma}_2$} &\multicolumn{1}{r}{$\|\bm{\sigma}_1\|$}  \\ \cmidrule(lr){2-2} \cmidrule(lr){3-3} \cmidrule(lr){4-4} 
		\multirow{1}{*}{$\bm{i}_{p}$}    & $ 86.51\cos t -63.22\sin t  +86.51\cos 3t + 63.22\sin 3t $ & $ -86.51\sin t - 63.22\cos t -86.51\sin 3t + 63.22\cos 3t $    & 107.15\\ [0.2cm]	
		\multirow{1}{*}{$\bm{i}_{q}$}    & $ 26.62\cos t +6.65\sin t  -26.62\cos 3t + 6.65\sin 3t $ & $ -26.62\sin t +6.65\sin t  +26.62\sin3t + 6.65\cos 3t $   & 27.44\\  [0.2cm]  
		\multirow{1}{*}{$\bm{i}_{F}$}  & $86.51\cos t + 86.51\cos 3t$ & $-86.51\sin t - 86.51\sin 3t$     & 86.51\\ [0.2cm]
		\multirow{1}{*}{$\bm{i}_{f}$}    & $ -63.22\sin t   + 63.22\sin 3t $  & $-63.22\cos t   + 63.22\cos 3t$     & 63.22\\ [0.2cm] 
		\multirow{1}{*}{$\bm{i}_{B}$}    & $6.65\sin t  + 6.65\sin 3t $ & $6.65\cos t  + 6.65\cos 3t$  &6.65\\  [0.2cm]
		\multirow{1}{*}{$\bm{i}_{b}$}   & $26.62\cos t   -26.62\cos 3t  $ & $-26.62\sin t   +26.62\sin 3t $  &26.62\\  [0.2cm]\cmidrule{2-3}
		\multirow{1}{*}{$\bm{i}$}       & $113.13\cos t - 56.57\sin t + 59.89\cos 3t + 69.87\sin 3t$ & $-113.13\sin t - 56.57\cos t - 59.89\sin 3t + 69.87\cos 3t$   & 110.61\\  \bottomrule
	\end{tabular}
	\vspace{+0.1cm}
	\caption{Current decomposition for circuit in Figure \ref{fig:RLC}.}
	\label{tab:RLC_currents_decomposition}
\end{table*}

Hence, it is possible to calculate the geometric power according to (\ref{eq:geom_power_time})

\begin{equation*}
\begin{aligned}
\bm{M}&=M_{p}+\bm{M}_{q}=\underbrace{24,470 + 17,882 \sin2t + 24,470\cos2t}_{M_p}\\
&+\underbrace{\left(1882+7530\sin2t + 1882\cos2t \right) \bm{\sigma}_{12}}_{\bm{M}_q}\\
\end{aligned}
\end{equation*}

The active power is  $P=\bar{M}_p/2=12,235$ W and the Budeanu reactive power $Q=\bar{\bm{M}_q}/2=941$ VAr. The current decomposition is derived according to the expression (\ref{eq:current_decomposition_initial}) and is reported in table \ref{tab:RLC_currents_decomposition}. Notice that the inverse of the voltage vector is

\begin{equation*}
\bm{u}^{-1}=\frac{\bm{u}}{\|\bm{u}\|^2}=\sqrt{2}\left(\frac{\cos t+ \cos 3t}{800\cos^2 t}\bm{\sigma}_1-\frac{\sin t+ \sin 3t}{800\cos^2 t}\bm{\sigma}_2\right)
\end{equation*}

\noindent and the RMS voltage is $\|\bm{u}\|=200$. As in illustration 1, the power factor is the unity once the circuit is compensated. 

\begin{equation*}
pf_{\text{final}}=\frac{\bar{M}_p}{\|\bm{M}_{\text{final}}\|}=\frac{\bar{M}_p}{\|\bm{u}\|\|\bm{i}\|}=\frac{24,470}{200 \cdot 86.51\sqrt{2}}=1
\end{equation*}

Note that maximum compensation should be carried out by active elements since the current $\bm{i}_f \neq 0$.

\section{Conclusion}

The instantaneous reactive theory IRP (and its CV version) have been a useful mathematical tool for the compensation of polyphase systems. It enables a decomposition of the current according to clear engineering terms: reduction of the source current without energy storage. However, its mathematical formulation is incomplete as it is not able to handle single-phase systems and cannot use averaged quantities. This leads to the generation of currents containing harmonics and asymmetrical components, even in systems supplied by sinusoidal, symmetrical voltages and with passive linear loads. In contrast, GAPoT theory addresses these challenges through the use of geometric algebra and Hilbert's transform. By defining the geometric apparent power as the product of  voltage and current vectors, a robust and compact formulation is achieved which captures the multi-component nature of power systems. The current decomposition can be performed in a natural and straightforward way, providing results that are in line with the physical intuition of the problem. Likewise, IRP theory is shown as a particular case of GAPoT, in which only instantaneous compensation is considered without averaging. 

\section*{Acknowledgment}

This research has been supported by the Ministry of Science, Innovation and Universities at the University of Almeria under the programme \textit{Proyectos de I+D de Generacion de Conocimiento} of the national programme for the generation of scientific and technological knowledge and strengthening of the R+D+I system with grant number PGC2018-098813-B-C33.

\bibliographystyle{IEEEtran}
\bibliography{mybib}

\begin{thebibliography}{10}
\providecommand{\url}[1]{#1}
\csname url@samestyle\endcsname
\providecommand{\newblock}{\relax}
\providecommand{\bibinfo}[2]{#2}
\providecommand{\BIBentrySTDinterwordspacing}{\spaceskip=0pt\relax}
\providecommand{\BIBentryALTinterwordstretchfactor}{4}
\providecommand{\BIBentryALTinterwordspacing}{\spaceskip=\fontdimen2\font plus
\BIBentryALTinterwordstretchfactor\fontdimen3\font minus
  \fontdimen4\font\relax}
\providecommand{\BIBforeignlanguage}[2]{{%
\expandafter\ifx\csname l@#1\endcsname\relax
\typeout{** WARNING: IEEEtran.bst: No hyphenation pattern has been}%
\typeout{** loaded for the language `#1'. Using the pattern for}%
\typeout{** the default language instead.}%
\else
\language=\csname l@#1\endcsname
\fi
#2}}
\providecommand{\BIBdecl}{\relax}
\BIBdecl

\bibitem{akagi1984instantaneous}
H.~Akagi, Y.~Kanazawa, and A.~Nabae, ``Instantaneous reactive power
  compensators comprising switching devices without energy storage
  components,'' \emph{IEEE Transactions on industry applications}, no.~3, pp.
  625--630, 1984.

\bibitem{depenbrock1993fbd}
M.~Depenbrock, ``The fbd-method, a generally applicable tool for analyzing
  power relations,'' \emph{IEEE Transactions on Power Systems}, vol.~8, no.~2,
  pp. 381--387, 1993.

\bibitem{AkagiInstantaneous}
H.~Akagi, E.~H. Watanabe, and M.~Aredes, \emph{Instantaneous Power Theory and
  Applications to Power Conditioning}.\hskip 1em plus 0.5em minus 0.4em\relax
  Wiley, 2007.

\bibitem{czarnecki2005instantaneous}
L.~S. Czarnecki, ``Instantaneous reactive power pq theory and power properties
  of three-phase systems,'' \emph{IEEE Transactions on Power Delivery},
  vol.~21, no.~1, pp. 362--367, 2005.

\bibitem{haley2015limitations}
P.~Haley, ``Limitations of cross vector generalized pq theory,'' in \emph{2015
  International School on Nonsinusoidal Currents and Compensation
  (ISNCC)}.\hskip 1em plus 0.5em minus 0.4em\relax IEEE, 2015, pp. 1--5.

\bibitem{de2005discussion}
F.~De~Leon and J.~Cohen, ``Discussion of" generalized theory of instantaneous
  reactive quantity for multiphase power system",'' \emph{IEEE Transactions on
  Power Delivery}, vol.~21, no.~1, pp. 540--541, 2005.

\bibitem{dai2004generalized}
X.~Dai, G.~Liu, and R.~Gretsch, ``Generalized theory of instantaneous reactive
  quantity for multiphase power system,'' \emph{IEEE Transactions on Power
  Delivery}, vol.~19, no.~3, pp. 965--972, 2004.

\bibitem{peng1996generalized}
F.~Z. Peng and J.-S. Lai, ``Generalized instantaneous reactive power theory for
  three-phase power systems,'' \emph{IEEE transactions on instrumentation and
  measurement}, vol.~45, no.~1, pp. 293--297, 1996.

\bibitem{salmeron2009instantaneous}
P.~Salmer{\'o}n and R.~Herrera, ``Instantaneous reactive power theory—a
  general approach to poly-phase systems,'' \emph{Electric Power Systems
  Research}, vol.~79, no.~9, pp. 1263--1270, 2009.

\bibitem{montoya2020geometric}
F.~G. Montoya, J.~Rold{\'a}n-P{\'e}rez, A.~Alcayde, F.~M. Arrabal-Campos, and
  R.~Banos, ``Geometric algebra power theory in time domain,'' \emph{arXiv
  preprint arXiv:2002.05458}, 2020.

\bibitem{montoya2020analysis}
F.~G. Montoya, R.~Ba{\~n}os, A.~Alcayde, F.~M. Arrabal-Campos, and E.~Viciana,
  ``Analysis of non-active power in non-sinusoidal circuits using geometric
  algebra,'' \emph{International Journal of Electrical Power \& Energy
  Systems}, vol. 116, p. 105541, 2020.

\bibitem{cafaro2017geometric}
C.~Cafaro, ``Geometric algebra and information geometry for quantum
  computational software,'' \emph{Physica A: Statistical Mechanics and its
  Applications}, vol. 470, pp. 154--196, 2017.

\bibitem{hestenes2012clifford}
D.~Hestenes and G.~Sobczyk, \emph{Clifford algebra to geometric calculus: a
  unified language for mathematics and physics}.\hskip 1em plus 0.5em minus
  0.4em\relax Springer Science \& Business Media, 2012, vol.~5.

\bibitem{bracewell1986fourier}
R.~N. Bracewell and R.~N. Bracewell, \emph{The Fourier transform and its
  applications}.\hskip 1em plus 0.5em minus 0.4em\relax McGraw-Hill New York,
  1986, vol. 31999.

\end{thebibliography}

\end{document}